J. Braun[1], I. Rahbari[2], G. Paniagua[1], P. Aye Addo[1],
J. Garicano-Mena[3], E Valero[1,3], S. Le Clainche[3]

[1]*Purdue University,* [2]*University of Southern California,* [3]*UPM Madrid*

Characterization of Shock-Separation Interaction Over Wavy-Shaped Geometries
Through Feature Analysis of the Experimental Data

Abstract

**A canonical wavy surface exposed to a Mach 2 flow is investigated through high-frequency Pressure Sensitive Paint (PSP), Kulite measurements, and shadowgraph imaging. The wavy surface features a compression and expansion region, two shock-boundary layer interactions, and two shock-separation regions. The unsteady characteristics of the wall pressure and shock angles are presented, demonstrating an increase in amplitude of the instabilities when traveling through the shock systems. Three-dimensional flow features, observed in PSP data, reveal a two-dimensional flow pattern. Higher-Order Dynamic Mode Decomposition and Spectral Proper Orthogonal Decomposition are implemented to dissect the different flow features, revealing several dominant low-frequency and medium-frequency phenomena. The separation region appears at frequencies with Strouhal numbers between 0.01 and 0.2, confirmed by the frequency content in the local pressure measurement using Kulites.**

**Highlights**

- Analysis of shock boundary layer and shock-separation interactions in wavy-shaped geometry
- 10 kHz shadowgraph for quantification of unsteady shock, shear layer, and separation regions

- Pressure Sensitive Paint (PSP) and Kulite measurements to quantify pressure unsteadiness on the walls and frequency spectra associated with the different features
- Higher order dynamic mode decomposition and spectral proper orthogonal decomposition

List of symbols

| | |
|---|---|
| P | Pressure [Pa] |
| T | Temperature [K] |
| $\delta_2$ | Momentum layer thickness [mm] |
| z | Axial direction [m] |
| $\varepsilon_{SVD}$ | tolerance employed in first SVD reconstruction step |
| $\varepsilon_{DMD}$ | tolerance employed in first SVD reconstruction step |
| d | number of time-lagged snapshots for HODMD |
| $n_s$ | amount of samples |
| f | Frequency [Hz] |
| A | Amplitude [-] |
| ***A*** | Matrix |
| $L_s$ | Separation length [m] |
| L | Length of the domain |
| $\Lambda$ | SPOD Eigenmode [-] |
| M | Mach number [-] |
| z | Axial direction [m] |
| y | Wall-normal direction [m] |
| **Subscripts** | |
| 0 | Total condition |
| 1 | Region 1 |
| 2 | Region 2 |
| **Abbreviations** | |

| SBLI | Shock boundary layer interaction |
|------|----------------------------------|
| C | Compression shock |
| SR | Separation region |
| AC | Unsteady component |
| SPOD | Spectral Proper Orthogonal Decomposition |
| HODMD | Higher Order Dynamic Mode Decomposition |

1. Introduction

The understanding of compression shock trains, expansion fans, and shock boundary layer interactions is essential to supersonic fluid machinery [1]. Over the past years, several concepts have been proposed, such as bladeless turbines, in which power is extracted through shock waves [2]. Axial supersonic turbomachinery detailed by Sousa et al. [3] have gained interest for flow conditioning downstream of pressure gain combustors and were evaluated for large amounts of unsteadiness by Liu et al. [4] and optimized by Mushtaq et al. for slender blade shapes [5]. Su et. al [6] performed large eddy simulations of oscillating inflow for supersonic passages [6]. Manfredi et al. recently proposed dedicated facilities for supersonic Organic Rankine nozzles with Mach numbers of around 1.8 and several shock boundary layer interactions [7]. Another option is supersonic radial inflow machinery [8], in which supersonic inflow enters the flow axially and exits radially with multiple compression and expansion waves. All these internal passages are subjected to multiple shock reflections, propagating downstream and possibly influencing the upstream shock structure. A precise understanding of the source and amplitude of the flow instabilities due to these shocks is required for accurate prediction of the structural and thermal loads on the device and the downstream flow. Hence, the experimental characterization of supersonic flows over canonical wavy surfaced geometries is essential for improving our understanding of supersonic internal passages, developing reduced-order models for designers, and improving our computational tools. A good way to analyze the combination of all these effects in an internal flow is studying a wavy geometry, described by Braun et al. [9] which consists of three sinusoidally shaped waves with an area increase to allow for continuous expansion of the flow. These waves give rise to compression shocks, shock boundary layer interactions, expansion fans and separation regions. The shock-separation flow features in high-speed internal flows over large amplitude wavy surfaces was not previously documented.

The primary physical phenomenon observed in these flow configurations is the Shock-boundary layer interaction (SBLI) which has been well-documented numerically by several authors, such as Tyson and Sandham [10], who conducted Direct Numerical Simulations for small amplitude-to-wavelength ratios, and Priebe and Martin who studied the flow over a canonical ramp using DNS [11]. The SBLI creates a strong, generally low-frequency, unsteady flow pattern caused by, among other effects, a consecutive attachment and detachment of the boundary layer [12]. However, this effect can be significantly modulated by various parameters; for example, see Clemens and Narayanswamy [13] for a review on how low-frequency phenomena are introduced in SBLIs. Several mechanisms are reported to be responsible for the modulations, either stemming from the upstream turbulent boundary layer, the downstream separated flow, or the intrinsically unstable nature of the SBLI system triggered by external perturbations. In this regard, Poggie and Smits [14] found that the upstream boundary layer could substantially influenced the intensity of the pressure fluctuations and the shock movement. This was also confirmed by Touber and Sandham [15], who analyzed several forcing functions on the incoming boundary layer. Through Direct Numerical Simulations, Wu and Martin [16] found that the separated flow region mainly determined the low-frequency shock motion, whereas the incoming upstream flow gave rise to the smaller-scale unsteadiness. Several researchers have studied the non-dimensional frequency content of traditional shock-boundary layer interactions. Kiya and Sasaki [17] demonstrated that the shear flapping fluctuations appeared at higher Strouhal numbers of approximately ~0.12. Deshpande and Poggie [18] reported the Strouhal number around 0.16 from a flat backward-facing step which triggered a separation shock and shear layer flapping. Using Dynamic Mode Decomposition of Wall-resolved LES calculations, Hu et al. [19] reported two distinct modes at Strouhal numbers of 0.02 and 0.2. The majority of these research concern *external* flow configuration that only exhibit one flow feature, SBLI. On the other hand, *very* low-frequency unsteadiness, with Strouhal numbers ranging between 0.02 and 0.05, has been associated with the reflected shock motions and separation bubble breathing [20]. Lui et al. investigated a supersonic stator passage through LES and found dominant peaks at Strouhal numbers of 0.045 and 0.13 on the suction side of the blade [21].

The present work aims to shed light on supersonic internal passages with multiple shock reflections and shock-separation features by applying *feature detection algorithms* to this specific problem. These algorithms are numerical tools designed to identify the underlying features of a high-dimensional dataset obtained from numerical or experimental observations. Such features - or modes- can describe the main physical processes happening in the dynamical system of interest and can be classified in frequency, energy, or any other variable of interest. To analyze the flow features, several different algorithms have been proposed. One of the most common is the Proper Orthogonal Decomposition (POD), which obtains a series of orthogonal modes only in space, and ranks them according to their energy contents. One of the alternative approaches to POD is the Dynamic Mode Decomposition (DMD) [22] which uses the Koopman linear operator (or the best linear approximation of the system) to calculate the Fourier-like expansions for the non-linear dynamics and the Spectral POD (SPOD) [23], which involves an additional temporal constraint to the classical POD and enables distinguishing the spatiotemporally coherent features. More recently, Le Clainche and Vega [24] proposed the Higher-Order DMD to extract the flow features using a series of time-lagged snapshots, resulting in a higher convergence and better accuracy of the modes and frequencies. Such tools are commonly used for supersonic flows. Liu et al. performed PSP on a ramp-induced shock and found modes similar to 2D canonical cases but also modes associated to three-dimensional wall effects SPOD and POD[25]. Furthermore, for Hoffman et al. show that not every mode at every single frequency has a clear physical interpretation, in their study through a cylinder-induced SBLI [26].

In this paper, the physical processes underlying the flow unsteadiness over the open-source wavy-shaped geometry of Braun et al. [9] exposed to a Mach 2 freestream are experimentally investigated using 10 kHz shadowgraph imaging, three-dimensional high-frequency Pressure Sensitive Paint, and pointwise high-frequency pressure measurements using Kulite sensors. This geometry creates shock-boundary layer and shock-separation features in the which propagate downstream, mimicking a supersonic expansion device. Non-dimensional frequency content at different locations are dissected and the amplitude of the characterized oscillations are measured, providing critical information for assessing the system's structural integrity and the aerodynamic response of the downstream components, downstream bladeless, or bladed turbine stages. The prepared datasets are finally analyzed with the algorithms SPOD and HODMD.

2. Flowfield

The test article's geometry, as shown in Figure 1a, consists of multiple hills and crests with wavy shapes that were determined following the process outlined in reference [27]. The flow exhibits subsequent events of compression and separation shocks created from the lower-end wall, resulting in shock boundary layer interactions on the upper-end wall. The test article's geometry is illustrated in Figure 1a, while the geometry placed inside the test section is shown in Figure 1b. The representative flow field over this geometry, at the freestream Mach number of 2.0, visualized through numerical Schlieren (which is defined as $\sqrt{\left(\frac{d\rho}{dy}\right)^2 + \left(\frac{d\rho}{dx}\right)^2}$), is depicted in Figure 1c. Five critical regions of interest are highlighted with red boxes in this figure. The compression ramp (C1) on the lower wall generates a shock boundary layer interaction (SBLI1) on the upper wall. The Strouhal numbers for the shock-boundary layer interactions is expected to lie between 0.02 and 0.04 [13], which is associated with the shock motion and breathing. Due to the acceleration in the rear part of the first wavy profile and the subsequent change in curvature to compress the flow in the second wavy contour, the first separation region appears ("separation region 1 (SR1)"). A shear layer separates the core supersonic flow from the separation region until the flow is recompressed further downstream at the onset of the second wavy surface.

The separation shock and the reflected shock from SBLI1 generate another SBLI interaction on the top wall, named SBLI2. The fifth region consists of a second separation region (SR2) on the second wavy contour's lower wall in the rear.

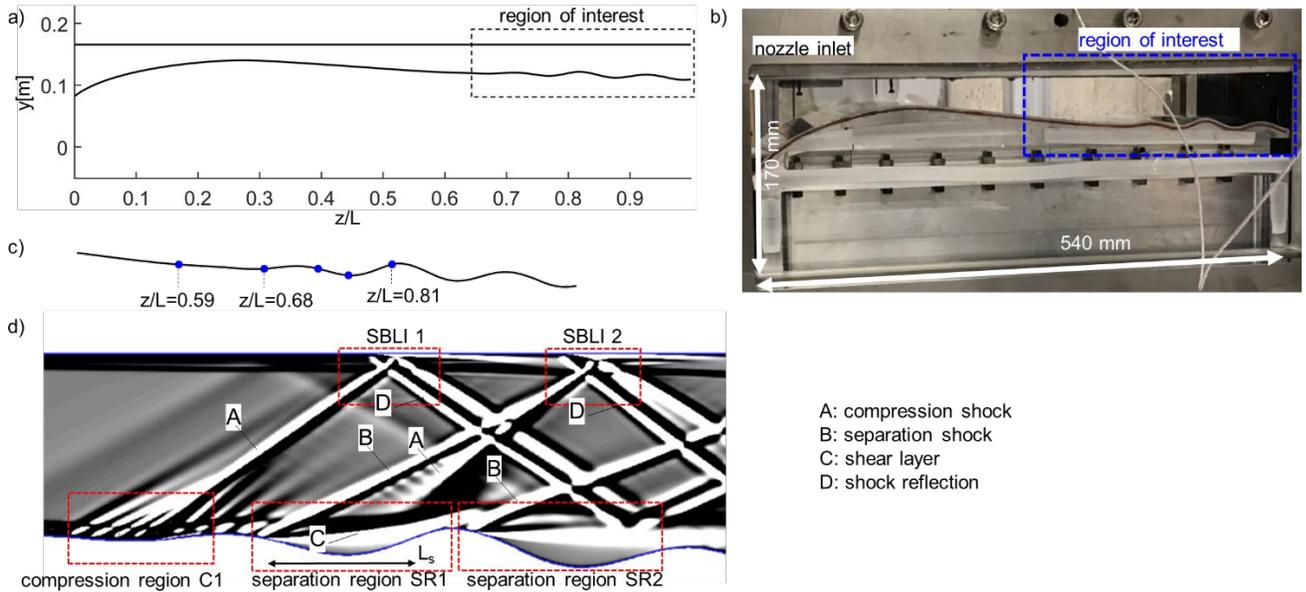

Figure 1 a) non-dimensional (z/L, where L=0.54m) coordinates of the test geometry, b) experimental test section, c) discrete measurement locations of the high frequency pressure measurements (blue dots), d) Numerical Schlieren of the wavy-shaped geometry in the region of interest marked in a)

Table 2 presents the inlet characteristics calculated based on a prior RANS simulation [18]. At 0.3m from the inlet, the mass-flow averaged Mach number is 2, the momentum thickness is approximately 0.5 mm, and the Reynolds number, based on this boundary layer thickness, is approximately $5.10^3$. The reference density and velocity numbers are 0.3 kg/m$^3$ and 505 m/s.

Table 1 Inlet characteristics of the wavy geometry

| Inlet BL | z/L [m] | M | δ2 [mm] | Re$_{δ2}$ |
|---|---|---|---|---|
|  | 0.55 | 2.01 | ~0.5 | ~$5.10^3$ |

Using the RANS simulations [18], the separation length (L$_s$) in each region of interest is calculated based on the averaged skin friction and presented in Table 2. The expected range of the "Strouhal number" in each region is taken from the literature discussed in the previous section and the corresponding frequency range, estimated according to these two parameters, is reported in Table 2. From this prediction, the dominant frequency of the first SBLI should lie between 1 and 3 kHz (St of 0.02-0.04), and the shear layer zone should fluctuate between 0.3 and 2.3 kHz (St of 0.02-0.2).

Table 2 Estimation of the dominant frequencies based on the characteristic length

|  | L$_s$ [mm] | M [-] | St | f [kHz] |
|---|---|---|---|---|
| **Compression 1** | 0 | 2 | 0.02 | - |
| **SBLI 1** | 7 | 2.08 | 0.02-0.04 | 1.15-2.9 |
| **SR 1** | 36 | 2.16 | 0.02-0.2 | 0.3-2.3 |
| **SBLI 2** | 11 | 2.04 | 0.02-0.04 | 0.7-1.8 |
| **SR 2** | 43 | 2.1 | 0.02-0.2 | 0.25-1.9 |

The state of the flow entering the test section has a substantial impact on the flow field within the regions of interest. The appendix shows hot wire experiments revealing an inlet turbulence intensity of approximately 4-6% [28].

3. Experimental setup

The experiments are performed at the Purdue Experimental Turbine Aerothermal Lab (PETAL) in the Linear Experimental Aerothermal Facility (LEAF). A detailed description of the facility is provided by Paniagua et al. [29]. Upstream of the test section, dry air at the total pressures of 2 bar and total temperature of 290K is discharged from a 10" fast-acting butterfly valve. The test section is under vacuum at 16kPa, achieved using a Dekker vacuum pump.

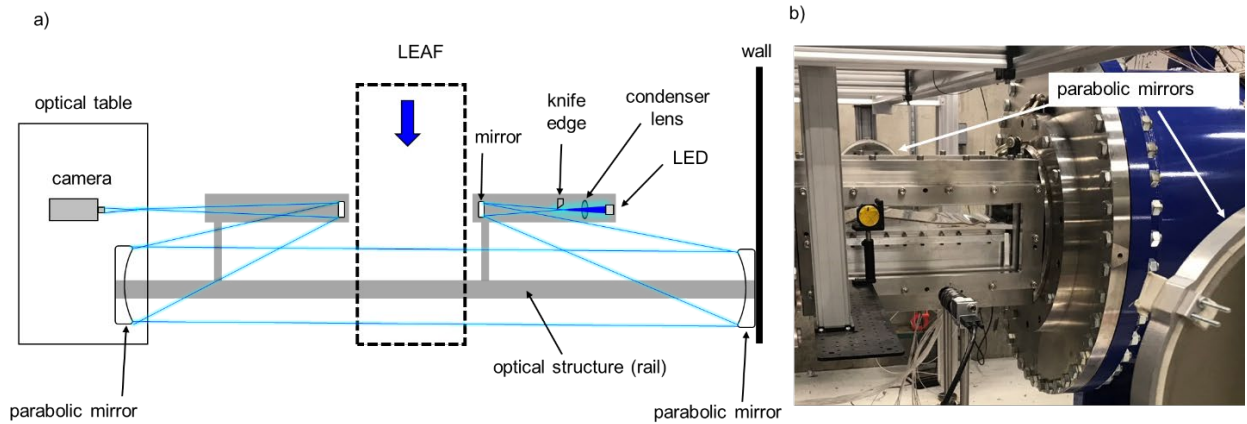

Figure 2 a) Schematic of the shadowgraph setup, b) Shadowgraph setup

Figure 2a shows the schematic view of the shadowgraph imaging setup with mirrors and the setup in the wind tunnel itself (Figure 2b). The shadowgraph system is of a Z-type configuration [30]. The camera is an SAZ Fastcam with the 10 kHz acquisition frame rate set. Fast response Kulite sensors (type XCE-062-50A Ultra-Miniature High-Temperature Pressure Probe), with a frequency response up to 120 kHz, are used for the pressure measurements. The Kulites are connected to a Precision Filter Instruments conditioning box (Model 101777 Triple Output Kulite Conditioner). The low-pass filter has a cut-off frequency of 322 Hz at -3 dB, and the band-pass filter has a frequency range of 200 Hz to 160 kHz. The sampling frequency for the Kulites is set to 512 kHz to ensure no aliasing is present. The Kulites are calibrated in the pressure range between

~ 10 kPa and the atmospheric pressure. Kulites are only mounted on the wavy shaped geometry, hence, the analysis of the unsteady flow features is focused mostly on the wavy shaped wall.

Pressure Sensitive Paint (PSP) is employed to obtain the pressure distribution over the wavy surface. Concerning the PSP, a SAZ Photron CMOS camera is sampled at 200 kHz to capture the time-resolved luminescence decay at a cropped resolution of 80×172 pixels. The camera is synchronized to a quasi-continuous pulse burst mode Nd: YAG laser operated at 10 kHz for a duration of 10 ms. Details on the calibration through pixel binning and fitting of an exponential decay function are provided by Aye-Addo et al. [31] as well as appendix A3. The acquisition frequencies and duration for the different measurement techniques are tabulated in Table 3.

Table 3 List of measurement techniques

| Measurement device | Acquisition frequency | duration |
|---|---|---|
| Scanivalve pressure sensors | 400 Hz | ~seconds |
| Kulites | 512 kHz | ~seconds |
| Shadowgraph | 10 kHz | 100 ms |
| PSP | 10 kHz | 10 ms |

4. Results

1 Wall pressure data

Figure 3a plots the high frequency Kulite pressure measurements at different locations along the test section during a typical experiment. The overall duration of one experiment is around 10 seconds. At the beginning, after valve opening, the bypass valve remains open and closes after 1 second where the test section becomes fully pressurized. A zoom on the startup is plotted in Figure 3b, for the first second, in which a gradual rise is observed in the upstream pressure ($z/L = 0.129$). The flow settles to supersonic conditions after around 0.4 seconds, which is corroborated by the shadowgraph images. The separation bubble appears in all instances and is pushed downstream when the supersonic flow is established. Kulites are only mounted on the wavy shaped geometry, hence, the analysis of the unsteady flow features is focused mostly on the wavy shaped wall.

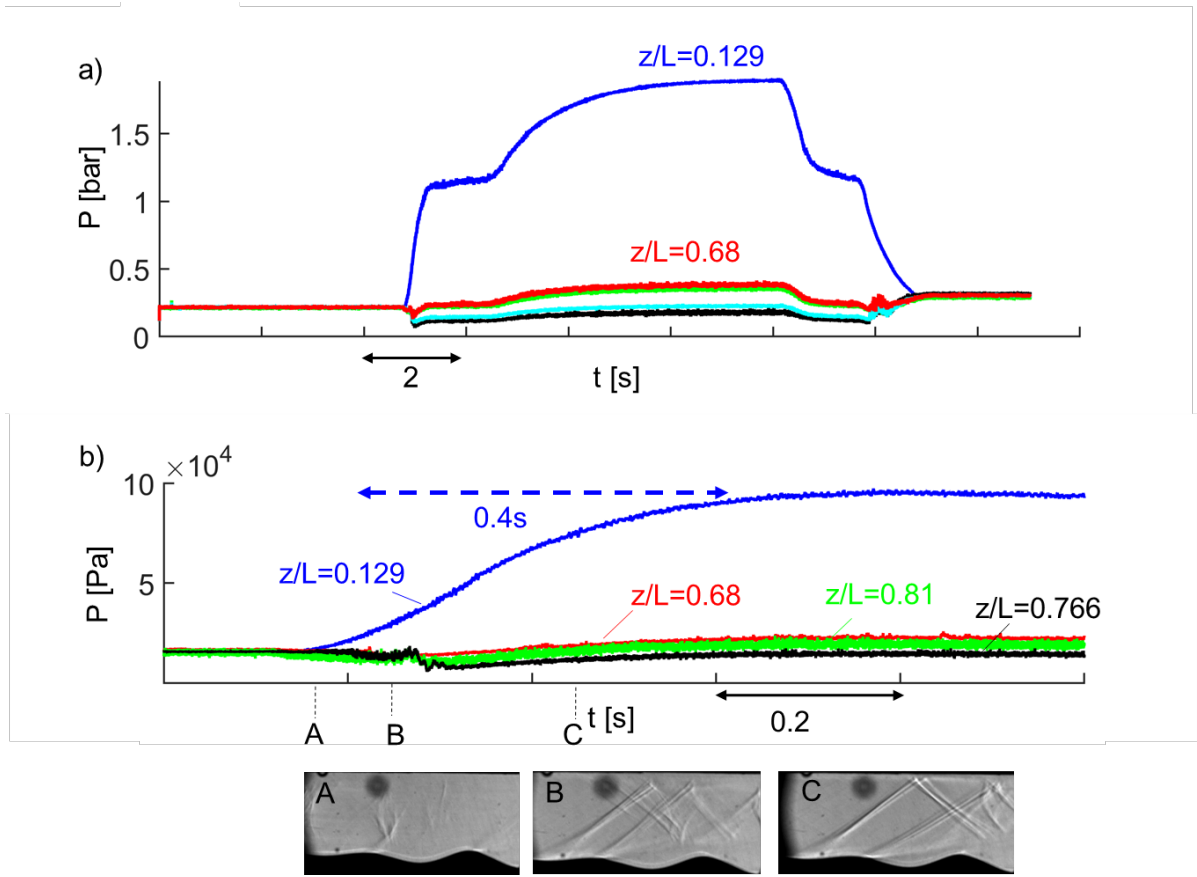

Figure 3 a) Pressure measurements at different locations along the test section during a typical experiment using Kulites, b) zoom on the start up process, c) isentropic Mach number as well as shadowgraph imaging at three instances

Figure 4 plots the average, along with the minima and maxima of three measurement techniques sampled during three different tests. This includes low frequency measurements using Scanivalves sensors (sampled at 400 Hz), high-frequency pressure measurements using Kulite miniature pressure transducers (sampled at 512 kHz), and the pressure sensitive paint (sampled at 10 kHz), as tabulated in Table 3. The pressure ratio is chosen to visualize the differences among the measurement techniques.

The Location of the separation shock agrees well with RANS prediction. Time-averaged pressure measurements using the Scanivalves and PSP shows the first separation shock's location between 0.722 and 0.725 and RANS predicts this shock to appear at z/L=0.728. The Kulite placed at this location measures high oscillations, due to the separation shock. The same applies to the second separation shock, at z/L=0.82. The overshoot in pressure ratio is larger in experiments compared to the RANS-predicted values which can be attributed to a misprediction in the separation point.

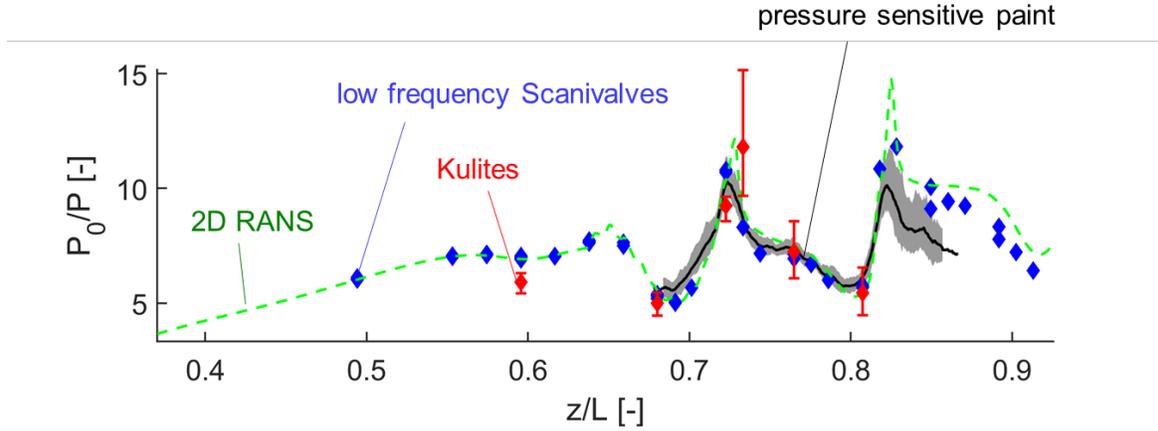

Figure 4 comparison between the low frequency pressure measurement (using Scanivalves scanners), high frequency pressure measurements (via Kulite transducers) and the pressure sensitive paint.

2 Spatio-temporal signature

This section aims to visualize the main flow structures and characterize their unsteadiness. Figure 5 illustrates a sequence of shadowgraph images during 12 milliseconds at a one-millisecond intervals. The readers are referred to the supplementary materials for the animation. From this set of shadowgraph images, the compression shock (C1) appears steady. However, the separation shock (indicated as number 1), the two shock boundary layer interactions on the top wall (SBLI1 and SBLI2), and the second separation region (indicated by number 3) have a higher signature of unsteadiness. The second shock boundary layer interaction (SBLI2) additionally exhibits an axial motion (represented by label 4 in Figure 5). These features can be better observed in the movie provided in the supplementary materials.

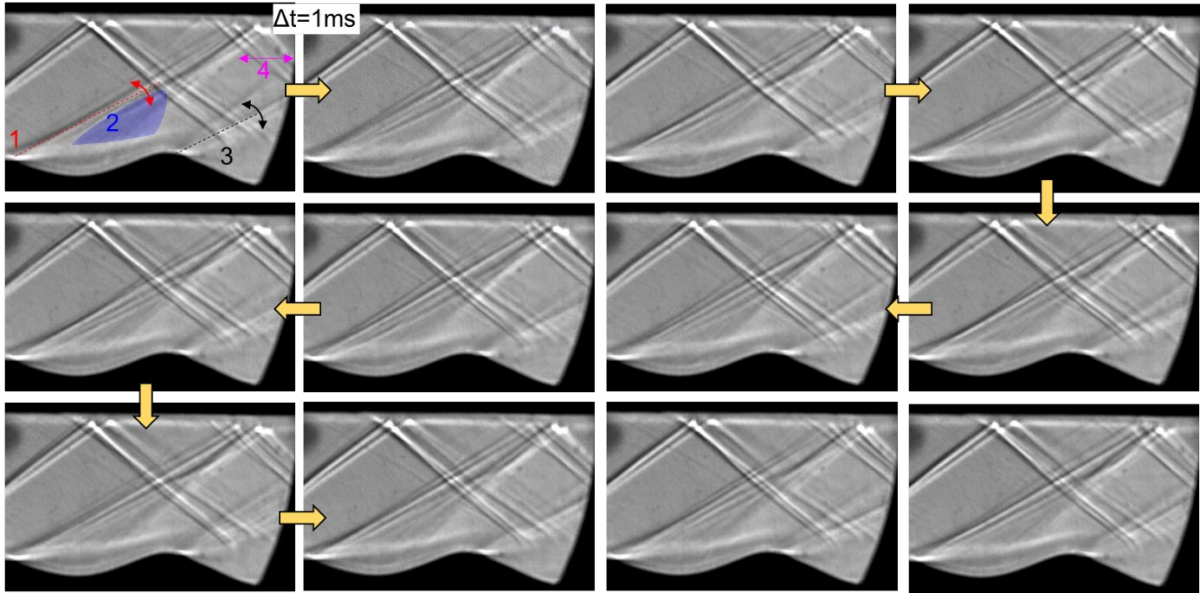

Figure 5 Shadowgraph imaging of the unsteady flow features during the steady operation, focused on the separation shock (1), the recompression zone (2), the second separation shock (3), and the second shock boundary layer interaction on the top wall (4)

Figure 6a depicts the shock angle fluctuations tracked using the high-frequency shadowgraph data. The grayscale images are binarized into 0's and 1's using a preset threshold of approximately 0.3. This technique, which is similar to the approach described by Tomasoni et al.[32], allows for isolating different shock structure in the flow. The shock location is defined by the point at which the value jumps from 0 to 1. These intersection points are collected and used for linear interpolation to calculate the shock angle. If the R-squared value is below 0.96, the angle is considered unreliable and discarded. The average angle of the first compression shock is around 34.6 degrees and, as evident from Figure 5 and discussed earlier, exhibits negligible fluctuations. The first separation shock, shown in red, originating from the separation point, has an oscillating behavior with a mean of 26.7 degrees with fluctuations up to 1.6 degrees. The shock angle of the first shock reflection on the upper wall (SBLI 1, magenta line) displays a periodic oscillation whereas the separation shock (red line) features an irregular, more aperiodic, behavior. To visualize the oscillations, Figure 6b plots the calculated angles on top of the original shadowgraph image.

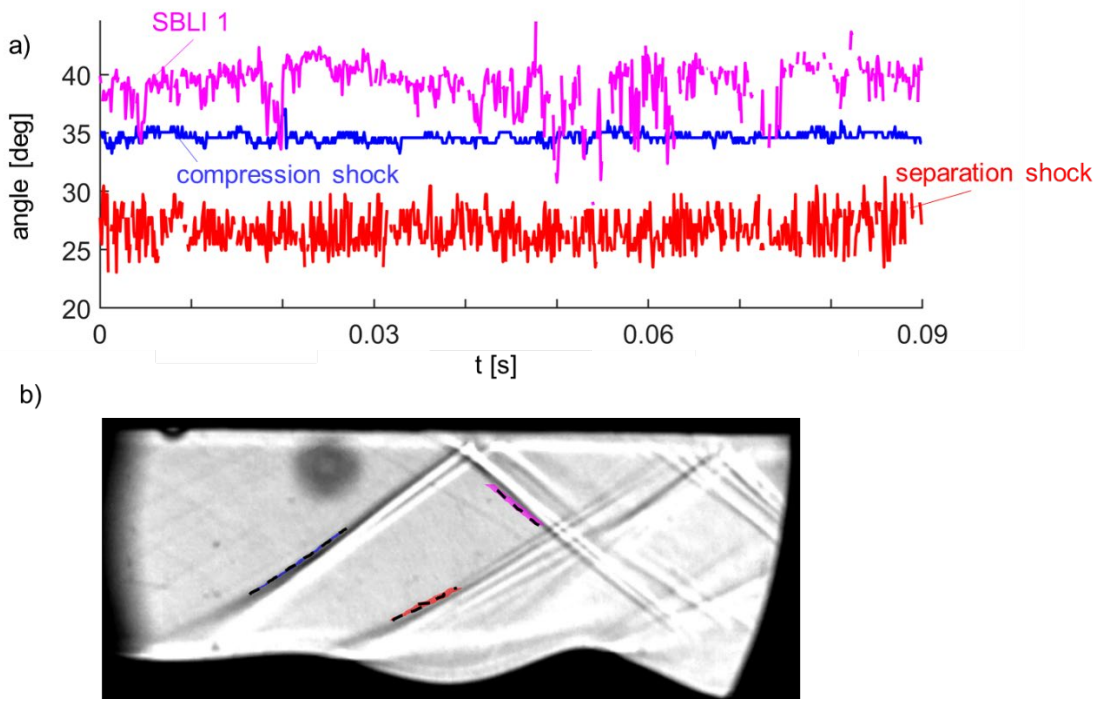

Figure 6 a) Mean flow angles as a function of time (blue= compression shock, red = separation shock, magenta = shock reflection) and b) Instantaneous shadowgraph with the flow angles variations

Table 4 tabulates the shock angle fluctuations. All angles are measured with respect to the axial direction. The compression shock angle measured using the high frequency shadowgraph agrees within 0.3 degrees of the results predicted by the 2D RANS calculations. SBLI1 angle is within 5 degrees of the RANS calculations, possibly owing to a slight misalignment in the shadowgraph experiments. A mismatch between the 2D RANS and the separation shock is also found, which is caused by a misprediction of the separation shock in the 2D RANS [9].

Table 4 Comparison of the shock angles and their standard deviation

| Angle name | Shadowgraph | 2D RANS |
| --- | --- | --- |
| Compression shock | 34.6±0.4 | 34.3 |
| First separation shock | 26.7±1.47 | 26.0 |
| SBLI 1 | 39.1±2 | 34.6 |

To quantify the three-dimensionality of this problem, Figure 7a plots the three-dimensional spatio-temporal variation of pressure from the PSP on the wavy wall (a top view) with a focus on the first compression, expansion, and separation region. The streamwise direction is defined by the coordinate "z", and the transversal direction by the coordinate "x". The x=0.115m location demarks the half-width of the test section. Separation region 2 features small structures moving in the transversal direction, while larger structures appear in the areas where compression occurs (C1 and C2). Figure 7b plots the time-averaged standard deviation in the spanwise direction for three data sets. Three zones of interest are identified. The first one is the region of the separation shock in which a higher amount of three dimensionality is present, owing to small curvature shock curvatures, the second zone is the recompression region (C2) and the third zone highlights the second separation shock region that features the highest three-dimensionality. In the latter, a spanwise distortion of approx. 7% was measured. The space-time contour plots for one test is presented in Figure 7c, where the black lines mark the first (0.725 z/L) and second separation shock (0.83 z/L). A sharper jump is observed in the second shock with higher levels of unsteadiness downstream of SR2. To highlight the distortion, Figure 7d plots the variation in pressure in the space-time diagram; unsteady flow motion is already measured in the expansion region upstream of SR1 as well as in the expansion region leading up to SR2.

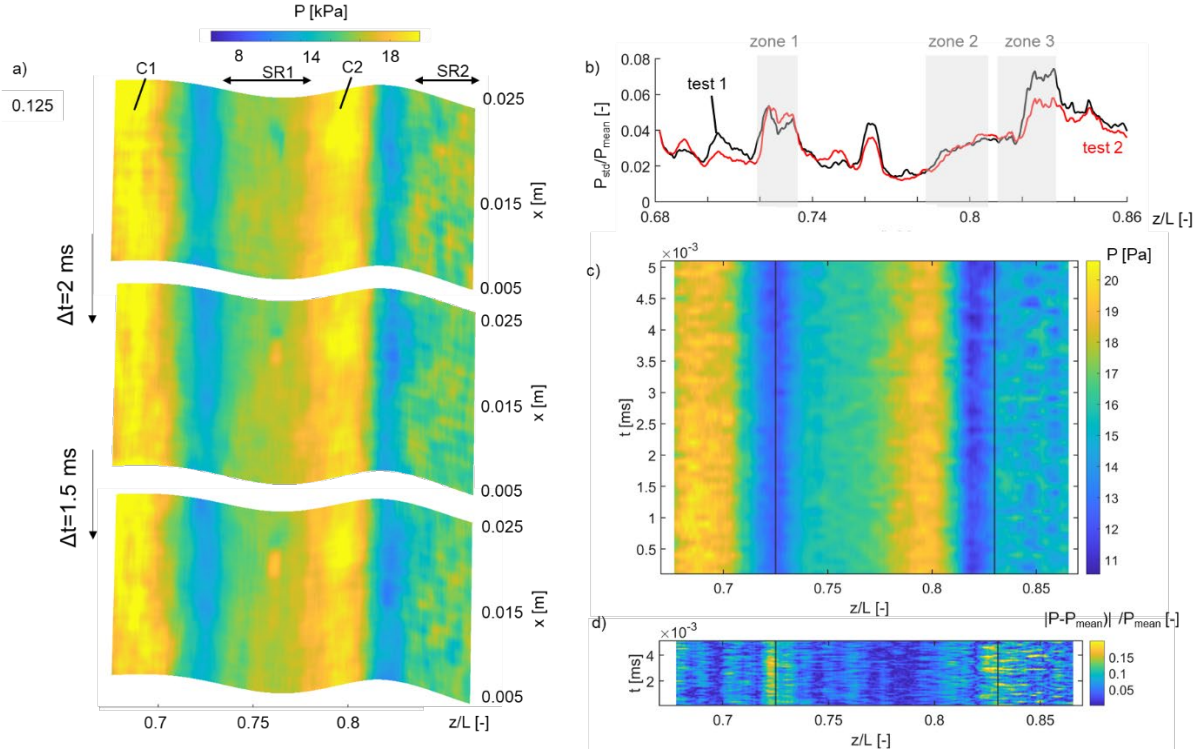

Figure 7 a) Instantaneous pressure field on the lower end wall with local acceleration and deceleration peaks at three different time steps (horizontal axis represent the axial and vertical axis shows the spanwise directions), b) dimensionless standard deviation of pressure versus the spanwise direction for two tests, c) Space-time diagram, d) Non-dimensional space-time diagram

These observations are also supported by pointwise pressure measurement using Kulite sensors. In Figure 8 a, the amount of unsteadiness is quantified by normalizing the standard deviation by the local mean pressure levels. The standard deviation of pressure fluctuations at the inlet is 0.3 % (not shown in the plot), and at the divergent part is 1.2%. This value jumps to 4.9% at the separation point and around 437% at in the recompression region C2 (at a location of $z/L = 0.81$). Normalizing the data against the upstream total pressure reveals a quasi-linear trend. The same procedure is followed for the PSP results which showed the overall lower pressure fluctuations, possibly owing to the limited samples ($n_s$ is around 100 shots for the PSP) and lower frequency resolution (10 kHz vs 512 kHz). In Figure 8 b, the probability density function of pressure fluctuation reveals an increase in unsteadiness when traveling downstream. A narrow Gaussian distribution at the inlet spreads progressively as one moves in the axial direction, suggesting that extreme events happen more frequently in downstream locations.

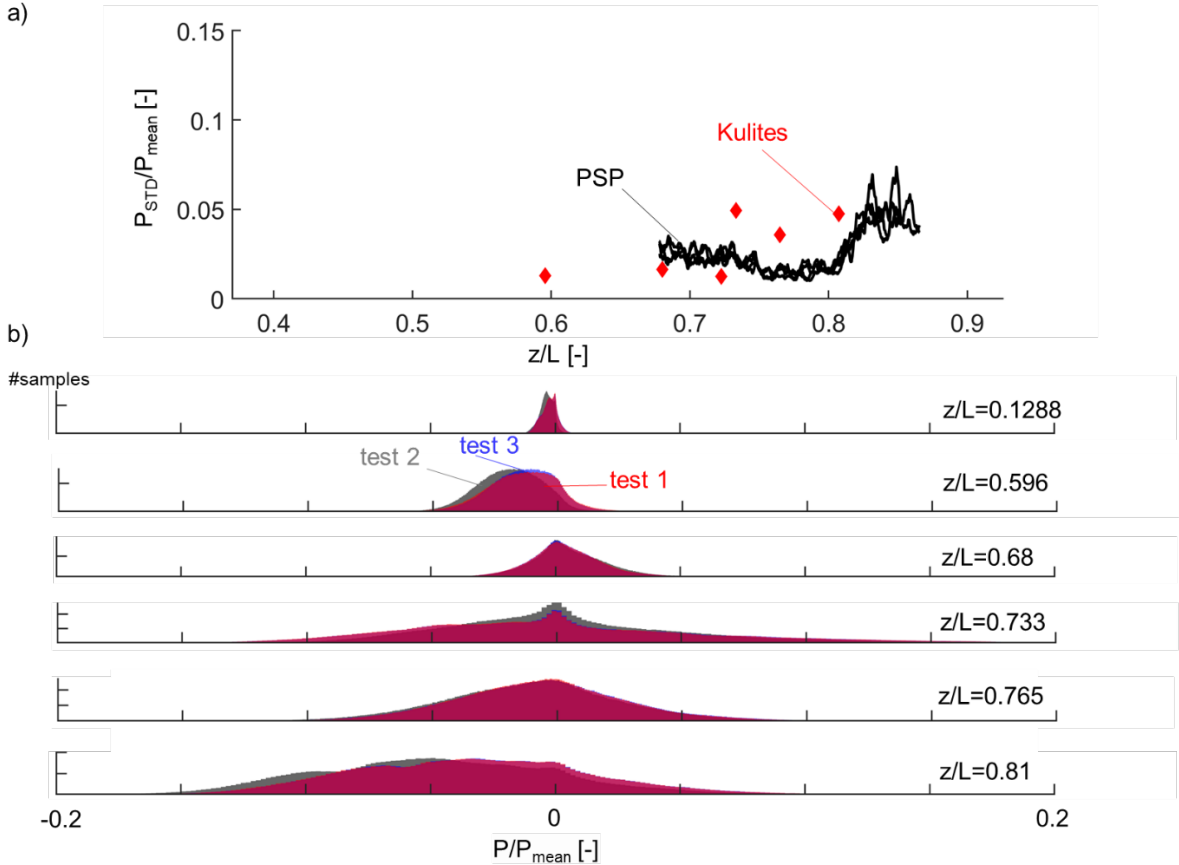

Figure 8 a) Standard deviation of pressure fluctuations during the steady-state operation measured using the Kulites and the PSP and b) Unsteadiness expressed as a function of the probability density function of pressure fluctuations measured using the Kulites at different axial locations

## 3 Frequency content on the wall

Figure 9 shows the Power Spectral Density (PSD) pre-multiplied by the frequency as a function of frequency and Strouhal number, calculated using the Kulite pressure measurements at several locations. The PSD was calculated based off the ratio of pressure by its mean value. As for the characteristic length, for calculating the Strouhal number, the separation length of the first separation region is used.

The amplitude of fluctuations in the inlet region ($z/L = 0.129$ in Figure 9a) is significantly smaller than in the divergent expanding section ($z/L = 0.596$). At the inlet, ($0.129\ z/L$), the low frequency oscillations are observed originating from the settling chamber. At the onset of the supersonic region, ($z/L = 0.596$), a high energy frequency range is observed ($St_{SR1} \sim 1$) which represents the large-scale turbulent eddies. This high frequency range is also observed in regions downstream and upstream of *compression ramps*, for instance reported by Priebe and Martin [11] and Poggie and Smits [14].

Figure 9b displays the results for the Kulite measurements upstream of the first separation shock, in the compression region (at z/L=0.68) and in the expansion region (at z/L=0.72). Low fluctuations are present at z/L=0.68, in the first compression region, at frequencies around 140 Hz. A second low frequency band is present as well (at frequencies of around 1400 Hz). We would expect an excitation of Strouhal of around 0.02 in that region triggered by the compression shock. These oscillations lie in the range of Strouhal numbers corresponding to the breathing of a possible separation bubble in the compression region [11]. Further downstream in the expansion region, these energy bands disappear and transition to a higher energy band at z/L=0.72.

Figure 9c displays the frequency content of pressure measurements in SR1 (z/L=0.735 downstream of the separation shock. When compared against the frequency spectrum of the pressure signal in the first shock region (at an axial location of z/L=0.68), it becomes clear that another peak has been added, due to the shear flapping motion which occurs at Strouhal numbers of 0.1 [33], in addition to the shock motion. Therefore, two distinct features are identified, one with a low Strouhal number between 0.0094 and 0.013 (~130Hz-180 Hz) and the other with a Strouhal number between 0.05 and 0.15 (~700Hz-2000 Hz). Further downstream (Figure 9c), this high energy frequency band gradually shifts to higher frequencies in the separation region (z/L 0.766). Furthermore, the Kulite measurements located in the separation region (at z/L=0.766, the red curve in Figure 9c) do not feature a distinct peak around the St~0.1. This was also observed by Deshpande and Poggie [33], who measured the wall spectra at multiple locations within the separated region. Further downstream, in the recompression region (z/L=0.81), a new frequency band appears, likely due to the shock recompression in that region. These spectra indicate that significant unsteadiness is found near the separation shocks, where the dominant frequencies upstream and downstream of the separation shock are within between Strouhal numbers of 0.02 and 0.2.

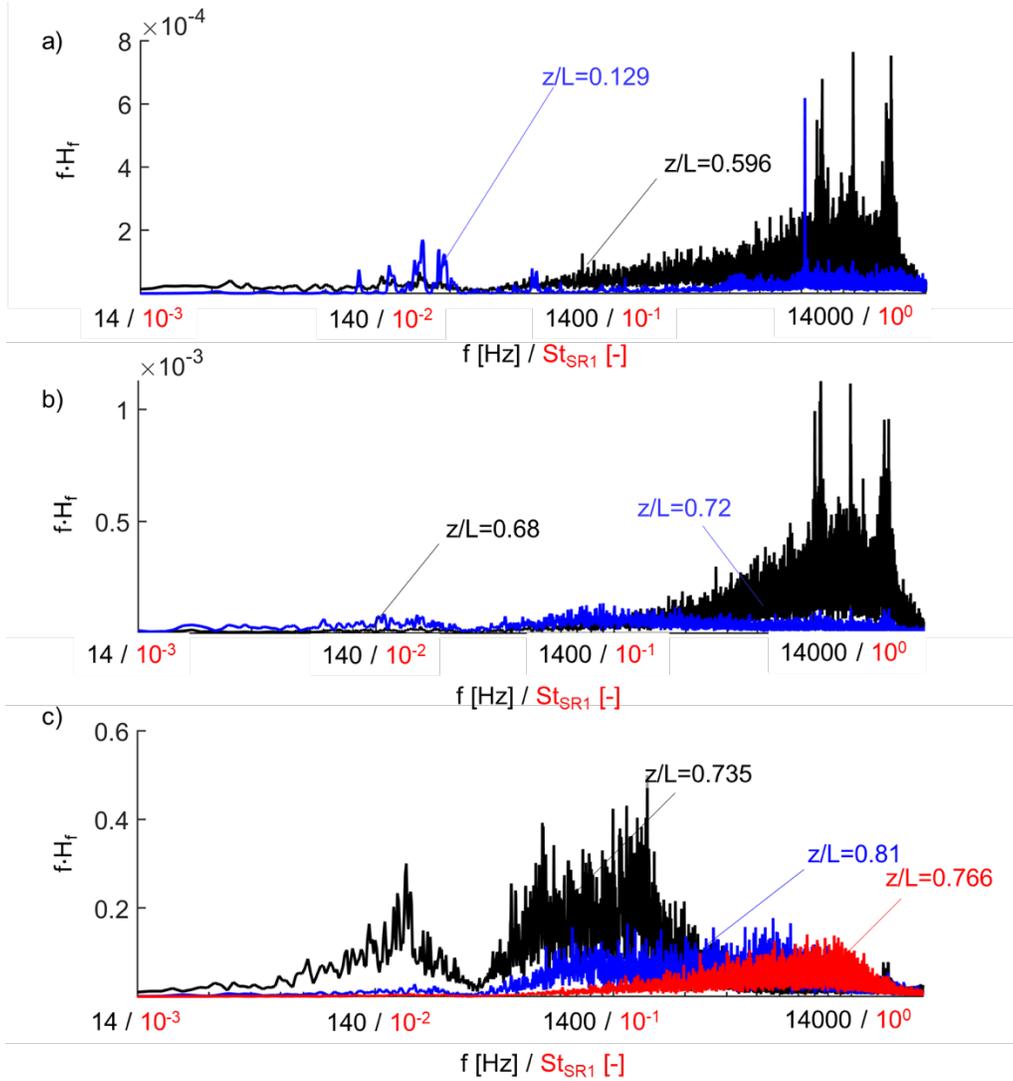

Figure 9 Pre-multiplied lower-end wall pressure spectra at the hub as a function of the Strouhal number (based on the separation length of separation region 1 for three different z/L (see annotated z/L locations in Figure 1) ($L_{SR1}=0.036$ and $u_{inf}=500$)

4 Spatio-temporal analysis through Spectral Proper Orthogonal Decomposition (SPOD)

A brief overview of the SPOD is given in appendix A1. For this analysis, 1200 snapshots ($n_s$) were used with windows of size 600 snapshots ($N_{FFT}$) and the overlap size of 500 images ($N_{overlap}$). For frequencies below 500Hz, Figure 10a shows the SPOD eigenvalues, pre-multiplied by their frequencies, for the first four dominant SPOD modes, as a function of frequency. The second and higher modes have significantly less energy than the first mode.-Figure 10b shows the contribution of each SPOD modes (7 SPOD modes in total) to the sum of eigenvalues, i.e. $\frac{\lambda_i}{\sum_{i=1}^{i=7}\lambda_i}$, at two selected frequencies (f=117 Hz in red triangles, f=1.65 kHz in black diamonds), and for the ensemble average of all frequencies (blue circles). The first mode has an average of 60% of all the energy content (but can be as high as 85% at 117Hz). This low-rank behavior is typical of SPOD spectra, as described by Towne et al. [23].

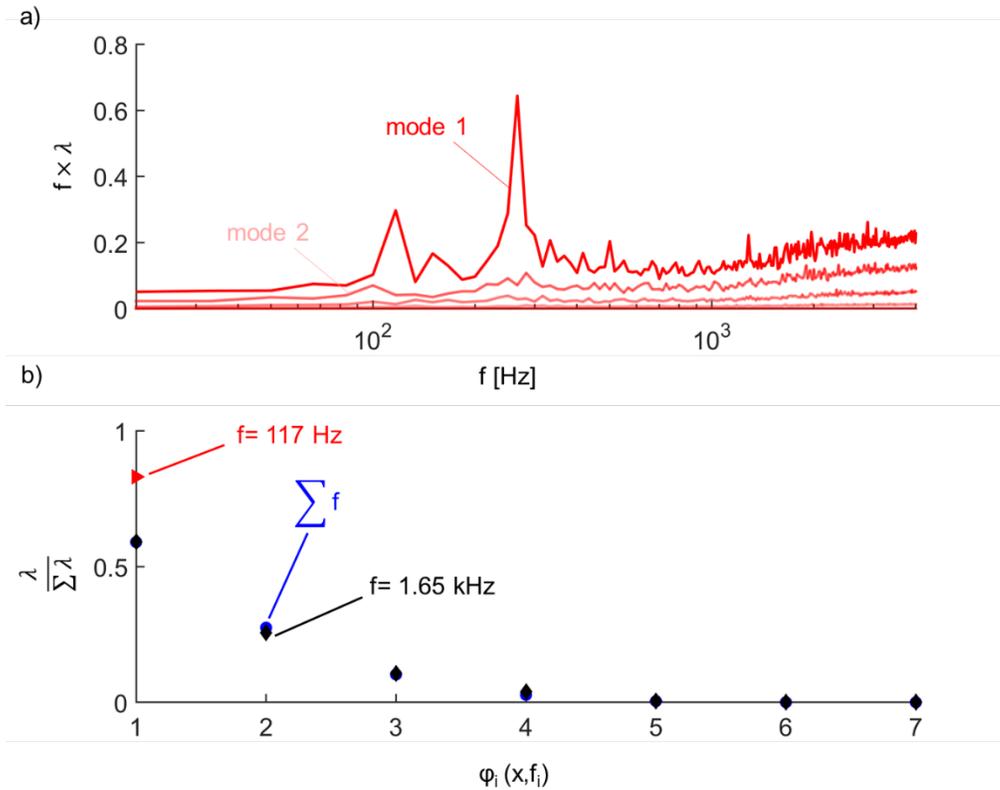

Figure 10 a) Pre-multiplied eigenvalue of the SPOD mode as a function of frequency and b) SPOD eigenvalues normalized by the total energy at two selected frequencies as a function of the SPOD mode number

Two frequency bands are considered hereafter: a lower-frequency range, up to 500 Hz and a higher frequency range, from 500 Hz to 4500 Hz. Figure 11a features the pre-multiplied SPOD eigenvalue spectrum of the first SPOD mode in the low-frequency band. Three peaks are observed: near 117Hz ($St_{SR1}$ 0.006), 250 Hz ($St_{SR1}$ 0.02), and 500 Hz ($St_{SR1}$ 0.04).

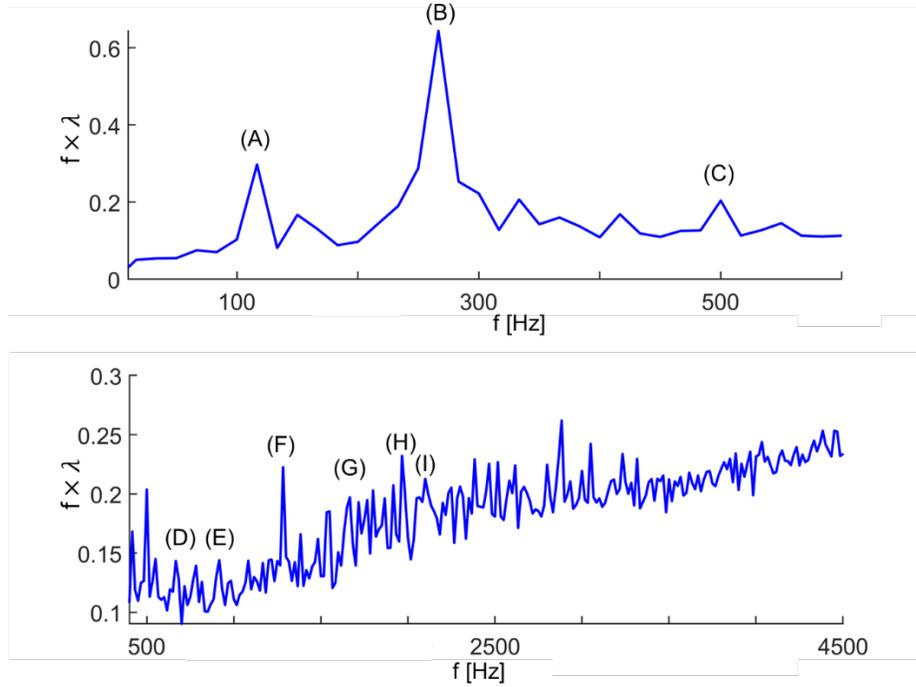

Figure 11 Spectrum of the pre-multiplied first SPOD mode from 10Hz to 500 Hz (a) and from 500 to 4500 Hz (b).

The mode shape corresponding to the frequency (A), i.e., the first SPOD mode shape at f=117Hz is analyzed in Figure 12. Figure 12a displays the first and second separation regions, and Figure 12b features a close-up in the first separation region. At this frequency, the flapping of the shear layer is observed, where an active area is located near the first and second separation regions. Moreover, the boundary layer upstream of SBLI1 and the incoming shock appear to have significant energy, possibly due to the convection of energy-containing flow structures from upstream, which was also identified in the PSD spectrum of the high-frequency pressure data at Z/L=0.12 (Figure 9a). The Strouhal numbers associated with this frequency, based on the second separation region size (see Table 2), is around 0.006.

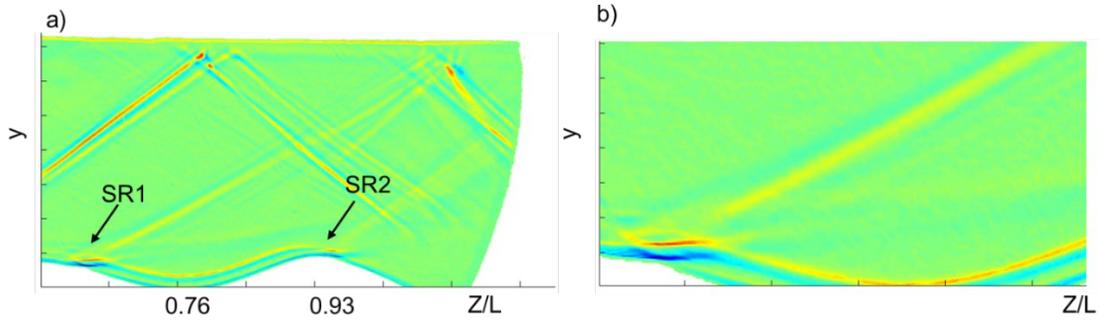

Figure 12 Fluctuation of the shear layers at 116 Hz: a) The entire flow field and b) Close-up of the first separation region (~$St_{SR1}$ 0.006)

The structure of the first SPOD mode at 266 Hertz (marked with label B in Figure 11) is depicted in Figure 13a. SBLI 1 and SBLI 2 appear to be the dominant features, while no active region is observed near the lower-end wall. Since the eigenvalue associated with the second SPOD mode at this frequency is about 1/3 of the first mode, its associated mode shape is shown in Figure 13b. Besides the slight differences in the mode shape on near the SBLI 1 and SBLI 2, the dominant flow features are similar to the first mode.

Figure 13c shows the first SPOD mode at 500Hz with a focus on the compression and first separation regions. The shear layer flapping and the interaction of the separation shock with the reflected shock (from SBLI1) are highlighted. When taking a closer look at the SPOD result (first mode at 500Hz) on both the first and second separation region (shown in Figure 13d), the second SBLI appears highly energetic.

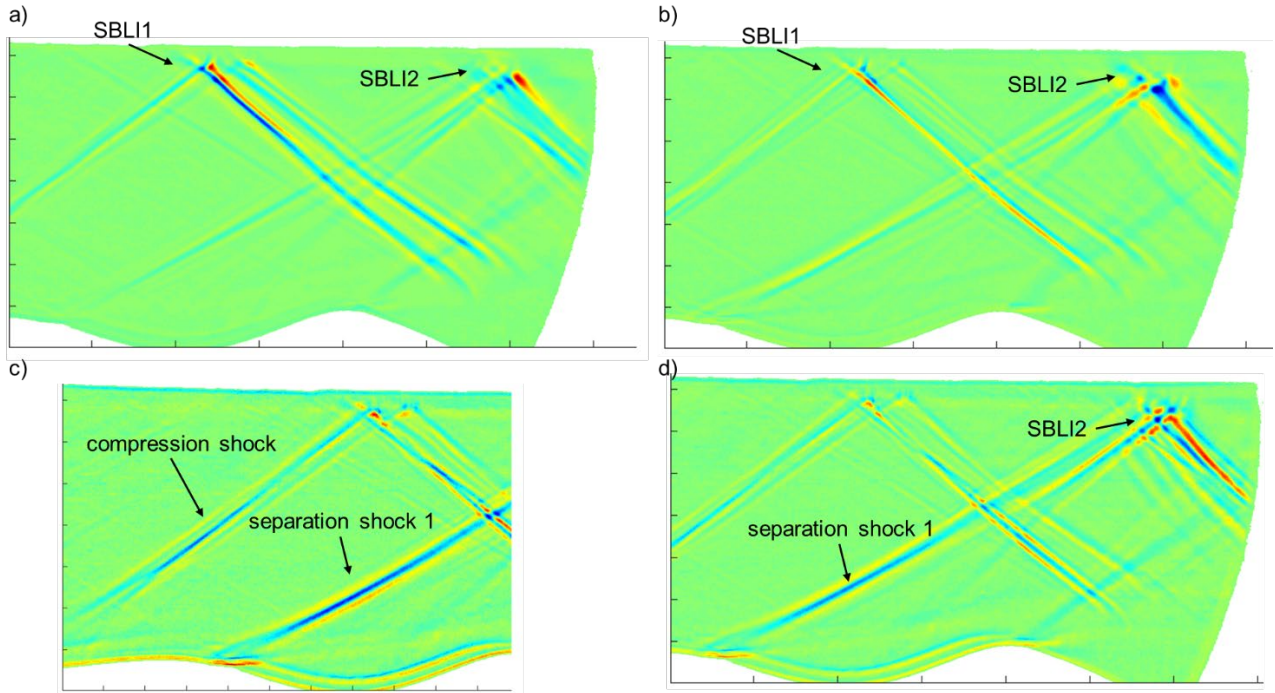

Figure 13 Fluctuations of SBLI 1 and SBLI 2: 266 Hz (~St 0.02 $_{SBLI1}$) a) the first mode and b) the second mode. The first mode at 500 hertz: c) zoom on the compression and the first separation region (~St 0.04 $_{SR1}$) and d) zoom on the first and the second separation region as well as SBLI2

In the higher frequency range, several other peaks in the SPOD spectrum are highlighted with letters (D) to (I) in Figure 14. The structure of the first mode at the frequency peaks (D), i.e., f=650 Hz, is shown in Figure 14a. At this frequency, corresponding to $St_{SBLI2}$, the second shock boundary layer (SBLI2) has the highest intensity. Figure 14b highlights the dominant mode's shape at frequency E, f=916 Hz, showing a mix of interactions. However, in contrast to the mode shape at f=650 Hz, the first separation shock responds to 916 Hz, corresponding to $St_{SBLI1}$ of 0.07. Figure 14c describes the structure of the first mode at peak (F), at 1280 Hz, in which the first separation shock is most active and for which pressure at the lower wall spectrum indicated higher levels of unsteadiness. At higher frequencies, SBLI dominates (e.g. Figure 14d at 2100 Hz).

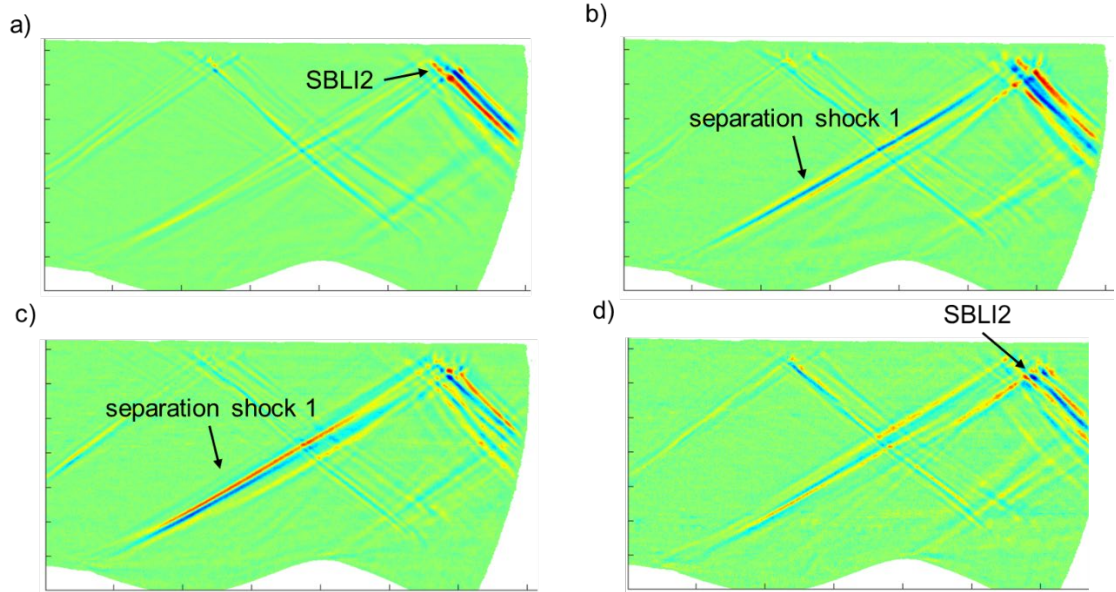

Figure 14 First SPOD mode shape at a) f=650 Hz, b) f=916 Hz, c) f=1280 Hz and d) f=2100 Hz

5 Robustness/Sensitivity to mode shape and frequency through higher order dynamic mode decomposition

To further evaluate the modes from section 4, HODMD is employed. A brief overview of the HODMD is given in appendix A1. SPOD and HODMD [34] are extensions of classical POD and DMD respectively. Both of them improve on the performance of the respective baseline method through a common strategy: exploiting data redundancy. SPOD leverages data redundancy in the spectral domain, by averaging spectra from shorter subsequences (in the spirit of the Welch method), whereas HODMD enacts data redundancy in the temporal domain through time–lagging of the data snapshots. Despite the similarities, the major differences from the underlying methods affect also SPOD and HODMD. Another difference between SPOD and HODMD derive from the fact that performing SPOD requires a single eigen factorization step, whereas HODMD uses two singular value decomposition steps and one eigen factorization step. This added complexity offers HODMD greater flexibility with respect to noise reduction and/or small-scale filtering of complex data (i.e. turbulence). This and other topics are discussed with more detail in reference [35]. The conclusion, following the findings in [35], is that SPOD and HODMD are powerful tools on their own, but they work better next to each other.

The HODMD method involves calibrating three parameters: the number of time-lagged snapshots (i.e., the temporal-redundancy level) d, and the tolerances $\varepsilon_{SVD}$ and $\varepsilon_{DMD}$ associated with the rank of the two truncated SVD approximations. Following [24], d~($n_s$/5,$n_s$/3), whereas $\varepsilon_{SVD}$ and $\varepsilon_{DMD}$ can be used to *filter out* the experimental noise, typically $\varepsilon_1$ and $\varepsilon_2$ are around ~($5\times10^{-3}$-$1\times10^{-2}$) [36]. Distinct features that are insensitive to variations of these three parameters are considered relevant features of the dataset, as opposed to numerical or noise-related "artifacts". Figure 15 shows the spectrum of the higher-order DMD. The spectrum indicates a series of modes with high energy, regardless of the filter employed. These modes appear at frequencies similar to the SPOD and are highlighted.

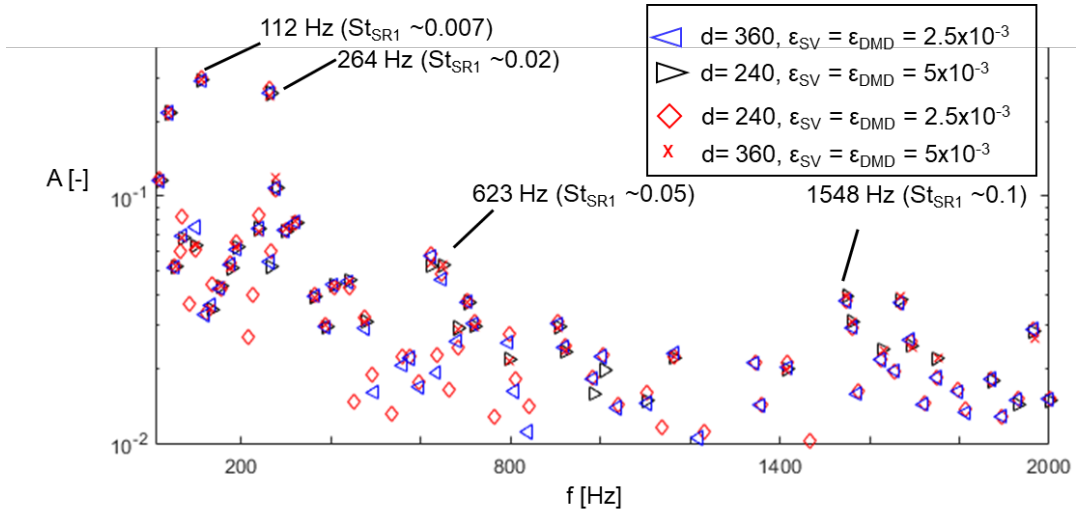

Figure 15 spectrum of the HODMD with several filters and 4 highlighted robust frequencies

Figure 16 illustrates the mode shapes associated with the frequencies carrying a large share of energy according to the HODMD spectrum in Figure 15. The Higher order DMD reveals that at lower frequencies, 40-265 Hz ($St_{SR1}$), all features are excited and could be partially excited through the turbulent inflow in the upstream nozzle, as evidenced by the turbulent inlet spectrum. In the 265 Hz range, all shock features are present (compression shock C1, first SBLI (SBLI1), SBLI2, as well as the separation shock. This is also apparent from the Kulites (see Figure). At higher frequencies, the first separation shock and compression shock are more energized, as shown in Figure 16b at 900 Hz in which A and B denote the first compression and separation shock. At 1100 Hz, other structures than shock structures are energized, such as the boundary layer in the recompression zone. This is also reflected in the Kulite data, at z/L=0.72 and 0.81 which contain energy at higher frequencies (1-14 kHz). Additionally, in all the investigated frequencies below 2 kHz, the shear flapping is visible (label A in Figure 16c).

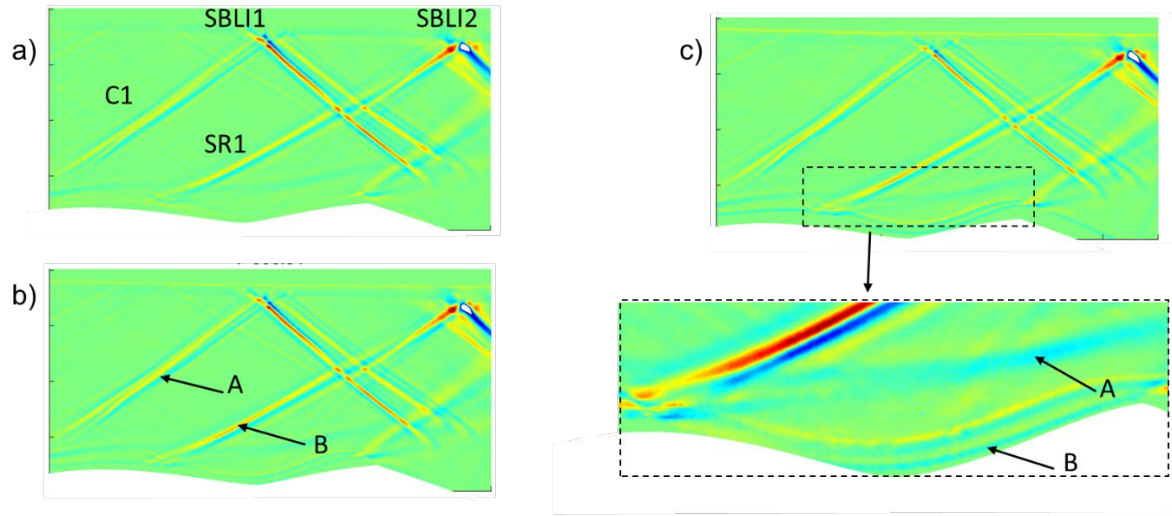

Figure 16 Spectra from the HODMD from low to high frequency, a) 265 Hz ($St_{SR1}$ ~0.015), b) 900 Hz ($St_{SR1}$ ~0.05), c) 1100Hz ($St_{SR1}$ ~0.1), and zoom on the first separation region (SR1)

5. Conclusions

Supersonic internal flow over a wavy surface is investigated through high-frequency pressure-sensitive paint, Kulite measurements, and shadowgraph. The channel features a compression shock, expansion waves, separation-reattachment shock on the bottom wall, and two shock boundary layer interactions on the top wall. The analysis identifies two main Strouhal regions between 0.01 and 0.1, linked to the breathing of the separation bubble typical for shock boundary layer interactions, and due to shear layer flapping in shock-separation region. Within the regions of accelerations, higher frequencie bands were found, with Strouhal numbers of around 1 were found. The physical structure of those frequencies is studied by two "state of the art" feature detection algorithms, namely Higher Order DMD and Spectral POD. The time series shadowgraph snapshots reveal several dominant low-frequency and medium-frequency phenomena of the separation regions and the two shock boundary layer interactions in the top wall.

Moreover, SPOD and HODMD algorithms can identify the frequency and the physical structures of the SBLI modes linked to lower Strouhal numbers. The modes show dominant shocks convected downstream and multiple reflections and interactions with the incoming boundary layer, creating a large recirculation region in the trough with the flapping of the detached shear layer.

The information provided by those algorithms is critical to identify the features and employ unsteady pressure and shock angle data for corrections of numerical models such as Reynolds Averaged Navier Stokes tools. Additionally, the precise knowledge of frequencies at which these features occur will help with structural design and future optimization of internal passages such as bladed and bladeless supersonic turbomachinery.

*Mech.*, vol. 915, 2021, doi: 10.1017/jfm.2021.95.

7. Acknowledgments

This work has been supported by SIMOPAIR (Project No. REF: RTI2018-097075-B-I00) funded by MCIN/AEI/10.13039/501100011033, and by the European Union's Horizon 2020 research and innovation program under the project FLOWCID, grant agreement No 101019137. The authors also acknowledge the support provided by Grant TED2021-129774B-C21 and by Grant PLEC2022-009235, funded by MCIN/AEI/10.13039/501100011033 and by the European Union "NextGenerationEU"/PRTR. J.G.M. acknowledges the support of "Programa Propio I+D+i 2020-Universidad Politécnica de Madrid

8. Appendix

A1 Feature detection algorithms

In this section, we briefly describe the two data-driven feature detection methods considered in this work, Spectral Proper Orthogonal Decomposition (SPOD, [23]) and Higher -Dynamic Mode Decomposition [24]. Both methods are derived from established techniques (Proper Orthogonal Decomposition (POD) and Dynamic Mode Decomposition (DMD), and provide improved results by exploiting data redundancy through windowing. For a detailed comparison of both methods, we refer to [37].

**Spectral Proper Orthogonal Decomposition**

To extract the spatio-temporal coherent structures within the flow field, we use the Spectral Proper Orthogonal Decomposition technique (SPOD). This method originates from a generic version of POD, initially proposed by Lumley [38], and seeks the most energetic modes that can vary in space and time. In the present work, we follow the Towne et al. implementation [39]. Where, vector $\boldsymbol{v}_i$ represents the flow quantities, here, the fluctuation of density gradient, at time instance $i$, $\boldsymbol{v}_i = \boldsymbol{v}_{i,\text{instantaneous}} - \overline{\boldsymbol{v}}$. An ensemble of $n_s$ frames are used to form the *data matrix*

$$V = [\boldsymbol{v}_1, \boldsymbol{v}_2, \dots, \boldsymbol{v}_{n_s}], \qquad Q \in \mathbb{R}^{I \times n_s}$$

where $I$ corresponds to the number of pixels per frame. Suppose we break up this matrix into $N_{\text{seg}}$ segments, each made of $N_{\text{FFT}}$ frames (spanning the time interval $T$) while sharing $N_{\text{overlap}}$ frames with the next segment:

$$\boldsymbol{V}_p = [\boldsymbol{v}_{(p-1) \times N_{\text{FFT}}+1}, \dots, \boldsymbol{v}_{p \times N_{\text{FFT}}}], \qquad p \in [1, N_{\text{seg}}]$$

or, for simplicity,

$$\boldsymbol{V}_p = [\boldsymbol{v}_1^p, \dots, \boldsymbol{v}_{N_{\text{FFT}}}^p], \qquad \text{or } [\boldsymbol{v}_k^p] \text{ where } k \in [1, N_{\text{FFT}}]$$

Taking Discrete Fourier Transform from each segment gives

$$\widehat{v}_k^p = \sum_{n=0}^{N_{\text{FFT}}-1} v_{n+1}^p \exp\left(-2\pi j \frac{kn}{N_{\text{FFT}}}\right).$$

A Hanning window is implemented to minimize spectral leakage due to imposing the periodicity window on the size of a segment. After this point, the analysis will continue separately at each frequency, $f_k = \left(k - \frac{N_{\text{FFT}}}{2}\right) \frac{N_{\text{FFT}}}{T}$, by creating the matrices

$$\widehat{V}_k = \left[\widehat{v}_k^1, \widehat{v}_k^2, \ldots, \widehat{v}_k^{N_{\text{seg}}}\right].$$

The Covariance kernel is then approximated as

$$C_k = \frac{1}{n_s - 1} \widehat{V}_k \widehat{V}_k^H$$

where the superscript $H$ denotes the Hermitian of the matrix. The eigenvalue decomposition of $C_k$ reads:

$$\boldsymbol{C_k W \Phi_k} = \boldsymbol{\Phi_k \Lambda_k}$$

where the columns of $\Phi_k$, i.e., $\phi_k^i$'s, are the eigenvectors and the diagonal elements of $\Lambda_k$, or $\lambda_k^i$'s, are the eigenvalues. The modes at each frequency are spatially orthogonal, subjected to the weight matrix, $\boldsymbol{W}$. In the present work, with the density gradient considered the flow quantity, the weight matrix is regarded as the identity.

**High-order Dynamic Mode Decomposition**

We apply the Dynamic Mode Decomposition to the time-resolved shadowgraph visualizations. A sequence of $n_s$ consecutive instantaneous shadowgraph images $v(t_k) \equiv \boldsymbol{v_k}$ acquired every $\Delta t_s$ time units are recast as real-valued vectors $\boldsymbol{v_k} \in \mathbb{R}^I$ with $I = n_x \times n_y$ where $n_x$ and $n_y$ are the horizontal and vertical image resolutions. The classical DMD method assumes these consecutive snapshots are linearly related, $\boldsymbol{v_{j+1}} = \boldsymbol{A v_j}$. The eigenvalues and eigenvector of the matrix operator $\boldsymbol{A} \in \mathbb{R}^{I \times I}$ expresses the dynamical information of the underlying physical system generating the images.

The numerical algorithm is summarized here for completeness; additional information can be found in [34] and [40].

Following the classical DMD method[41], the data matrices $V_1^{n_s-1} = (v_1, v_2, \ldots, v_{n_s-1}) \in \mathbb{R}^{I \times n_s-1}$ and $V_2^{n_s} = (v_2, \ldots, v_{n_s}) \in \mathbb{R}^{I \times n_s-1}$ are constructed such as:

$$\boxed{\boldsymbol{V_2^{n_s}} = \boldsymbol{A V_1^{n_s-1}}.} \quad (1)$$

Where the matrix $A$ is "a priori" unknown. To obtain the eigenspectrum of $A$, an economy-sized Singular Value Decomposition of $V_1^{n_s-1}$ is performed:

$$V_1^{n_s-1} = LSR^T \qquad (2)$$

where the superscript T denotes conjugate transpose. Matrix $S$ is a diagonal matrix whose entries $\sigma_i$, are the singular values. The left singular vectors – the columns of $L$ – can be related to the POD modes of the input data sequence. The reduced matrix, defined as $\tilde{A} = L^T A L = L^T A (V_1^{n_s-1} R S^{-1}) = L^T V_2^{n_s} R S^{-1}$, represents the projection of matrix $A$ on the $L$ space. After calculating the reduced matrix's eigenvalue decomposition, $\tilde{A} q_i = \mu_i q_i$, the DMD modes of the original matrix are found following $\phi_i = L q_i$ with the growth rates $\lambda_i = \log(\mu_i)/\Delta t_s$.

The Higher Order DMD algorithm substitutes the assumption on the linear between the consecutive snapshots with a more general d-lagged higher order Koopman assumption:

$$v_{k+d} = A_1 v_k + A_2 v_{k+1} + \cdots + A_d v_{k+d-1} \qquad (3)$$

Based on the SVD of the snapshot matrix $V_1^{n_s} = LSR^T$, we introduce the *chronos* matrix $\hat{V}_1^{n_s} \equiv \hat{S}\hat{R}^T \in \mathbb{R}^{r_1 \times n_s}$, where $r_1 \leqslant min(I, n_s)$. Equation (3) also holds for the *chronos* matrix. We rearrange these snapshots into *d* different matrices:

$$\hat{V}_j^{n_s-d+j} \in \mathbb{R}^{r_1 \times (n_s-d)}, j=1, \dots, d \qquad (6)$$

And stack them vertically, from j=1 to j=d, to create the block matrix:

$$\widetilde{V}_1^{n_s-d+1} = \begin{bmatrix} \hat{V}_1^{n_s-d+1} \\ \hat{V}_2^{n_s-d+2} \\ \dots \\ \hat{V}_{d-1}^{n_s-1} \\ \hat{V}_d^{n_s} \end{bmatrix} \in \mathbb{R}^{(d \times r_1) \times (n_s-d)} \qquad (7)$$

This block matrix is also factorized using SVD, $\widetilde{V}_1^{n_s-d+1} = \mathcal{L}\mathcal{S}\mathcal{R}^T$. Defining $\overline{V}_1^{n_s-d+1} \equiv \mathcal{S}\mathcal{R}^T \in \mathbb{R}^{r_2 \times (n_s-d)}$ and considering $r_2 \leqslant n_s - d$ offers a second opportunity to construct a reduced version of the database. At this stage, matrix $\overline{V}_1^{n_s-d+1}$ is split into matrices $\overline{V}_1^{n_s-d}$ and $\overline{V}_2^{n_s-d+1}$, and a linear relationship between them is imposed:

$$\overline{V}_2^{n_s-d+1} = \overline{A}\,\overline{V}_1^{n_s-d}. \qquad (8)$$

Given $\overline{V}_1^{n_s-d} = \overline{LSR}^T$, the matrix $\overline{A}$ is determined following

$$\overline{A} = \overline{V}_2^{n_s-d+1}\overline{R}\,\overline{S}^{-1}\overline{L}^T. \qquad (9)$$

and its eigenvalue decomposition, $\overline{AQ} = \overline{Q}M_\mu$, provides the reduced HODMD modes as the columns $\overline{q}_l$ of the matrix $\overline{Q}$, whereas the diagonal entries in $M_\mu$, $\mu_l$, determine the dynamics of the flow. $\lambda_l = \log(\mu_l)/\Delta t_s \equiv \alpha_l + i\omega_l$, where $\alpha_l$ indicates the amplification rate and $\omega_l$ represents the angular velocity of the mode $\overline{q}_l$.

A2 Hot wire measurements

Figure 17 displays the spectrum of velocity measured with a hot wire at three different distances away from the bottom surface for the same inlet condition during the commissioning of the wind tunnel at subsonic speeds (hotwire 1 within the freestream and hotwires 2 and 3 close to the end wall). At the range of Mach numbers studied here, inlet turbulence was around 4-6% [42] [43]. The hot wire that was employed was a mono-dimensional hot wire probe [44].

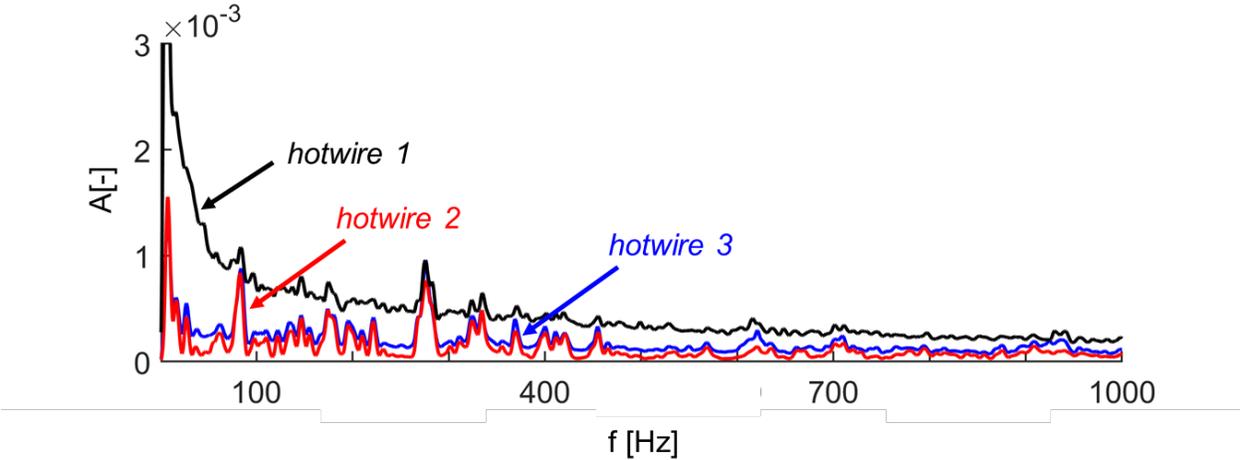

Figure 17 Inlet spectrum based on hotwire measurements from wind tunnel commissioning testing

A3 Pressure sensitive paint calibration

The PSP used in this experiment is a Platinum tetra-fluorenyl porphyrin (PtTFPP) luminophore with a porous polymer ceramic binder from Innovative Scientific Solutions, Inc. (ISSI). After applying layers of PSP to the test article with an airbrush spray gun, the PSP is calibrated under vacuum by controlling the test section pressure from 101.15 kPa to 16.3 kPa using a vacuum pump. Static pressure measurements with a Scanivalve DSA 3217 pressure scanner are used as a reference to calibrate the PSP. At each calibration point, a trigger signal from a Quantum Composer 9530 pulse generator is used to start the timing sequence of the experiment. A quasi-continuous burst mode Nd: YAG laser is pulsed at 10 kHz for a duration of 10 ms. Between each laser pulse, a Photron SAZ camera acquires 20 images at 200 kHz to capture the fluorescence decay within a time window of 100 micro-seconds. For each binning bundle of pixels, a single exponential decay function is used to model the luminescence decay and 2$^{nd}$ order polynomial coefficients are solved using a least squares fitting model that relates luminescent lifetime to pressure. Figure 18a shows the laser and camera mounting, and Figure 18b depicts the binder on the aluminum surface. In Figure 1c the combined binder and camera setup on the wavy surface are visualized. Figure 18d depicts the overall uncertainty which is around 4% for the lowest pressures.

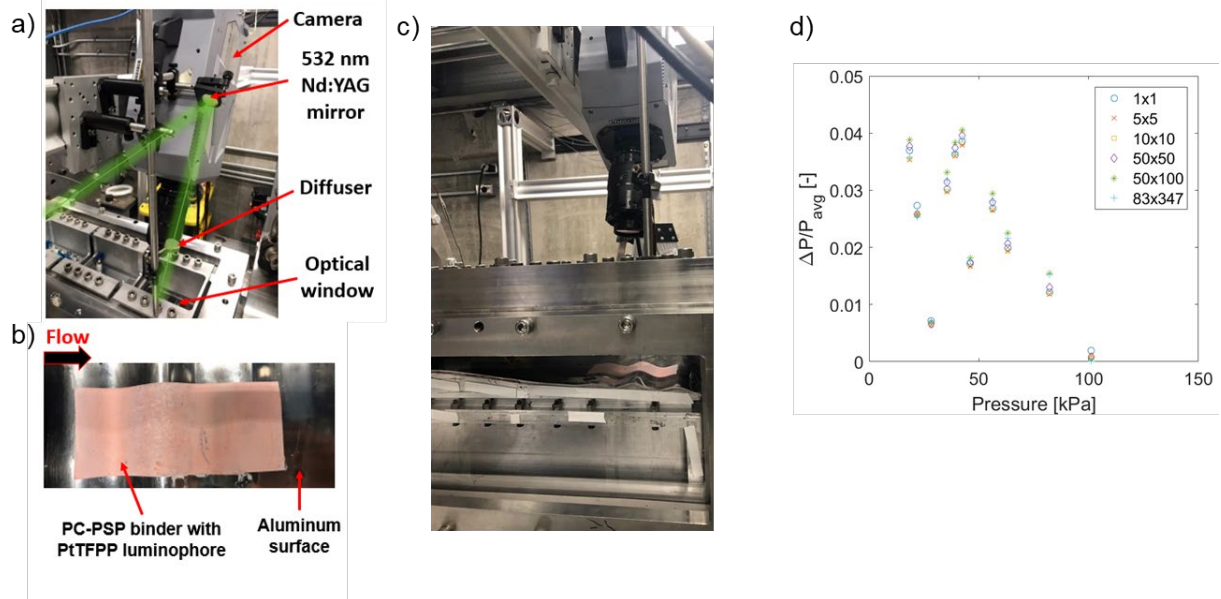

Figure 18 PSP Calibration: a) laser and camera mounting, b) binder on the wavy surface, c) viewing of the camera and the test article, c) uncertainty as a function of pressure for the PSP ,,

## A4 Geometry of the wavy surface

| # | Z[m] | Y[M] |
|---|---|---|
| 1 | 0.00000 | 0.0822 |
| 2 | 0.00001 | 0.0822 |
| 3 | 0.00003 | 0.0822 |
| 4 | 0.00005 | 0.0823 |
| 5 | 0.00007 | 0.0823 |
| 6 | 0.00010 | 0.0823 |
| 7 | 0.00012 | 0.0824 |
| 8 | 0.00015 | 0.0824 |
| 9 | 0.00019 | 0.0825 |
| 10 | 0.00022 | 0.0825 |
| 11 | 0.00026 | 0.0826 |
| 12 | 0.00031 | 0.0826 |
| 13 | 0.00036 | 0.0827 |
| 14 | 0.00042 | 0.0828 |
| 15 | 0.00048 | 0.0829 |
| 16 | 0.00054 | 0.0830 |
| 17 | 0.00062 | 0.0831 |
| 18 | 0.00070 | 0.0832 |
| 19 | 0.00079 | 0.0833 |
| 20 | 0.00089 | 0.0835 |
| 21 | 0.00100 | 0.0836 |
| 22 | 0.00112 | 0.0838 |
| 23 | 0.00125 | 0.0840 |
| 24 | 0.00140 | 0.0842 |
| 25 | 0.00155 | 0.0844 |
| 26 | 0.00173 | 0.0846 |
| 27 | 0.00192 | 0.0849 |
| 28 | 0.00213 | 0.0851 |
| 29 | 0.00236 | 0.0854 |
| 30 | 0.00261 | 0.0858 |
| 31 | 0.00289 | 0.0861 |
| 32 | 0.00319 | 0.0865 |
| 33 | 0.00352 | 0.0869 |
| 34 | 0.00389 | 0.0874 |
| 35 | 0.00428 | 0.0878 |
| 36 | 0.00472 | 0.0884 |
| 37 | 0.00520 | 0.0889 |
| 38 | 0.00573 | 0.0895 |
| 39 | 0.00631 | 0.0902 |
| 40 | 0.00695 | 0.0909 |
| 41 | 0.00764 | 0.0916 |
| 42 | 0.00841 | 0.0924 |
| 43 | 0.00925 | 0.0933 |
| 44 | 0.01018 | 0.0942 |
| 45 | 0.01119 | 0.0952 |
| 46 | 0.01231 | 0.0962 |
| 47 | 0.01354 | 0.0974 |
| 48 | 0.01489 | 0.0985 |
| 49 | 0.01637 | 0.0998 |
| 50 | 0.01800 | 0.1011 |
| 51 | 0.01979 | 0.1025 |
| 52 | 0.02176 | 0.1040 |
| 53 | 0.02392 | 0.1055 |
| 54 | 0.02629 | 0.1071 |
| 55 | 0.02889 | 0.1088 |
| 56 | 0.03174 | 0.1105 |
| 57 | 0.03486 | 0.1123 |
| 58 | 0.03829 | 0.1142 |
| 59 | 0.04203 | 0.1161 |
| 60 | 0.04612 | 0.1181 |
| 61 | 0.04967 | 0.1197 |
| 62 | 0.05333 | 0.1212 |
| 63 | 0.05702 | 0.1227 |
| 64 | 0.06073 | 0.1241 |
| 65 | 0.06448 | 0.1255 |
| 66 | 0.06824 | 0.1267 |
| 67 | 0.07203 | 0.1280 |
| 68 | 0.07583 | 0.1291 |
| 69 | 0.07966 | 0.1302 |
| 70 | 0.08349 | 0.1312 |
| 71 | 0.08735 | 0.1322 |
| 72 | 0.09121 | 0.1332 |
| 73 | 0.09509 | 0.1340 |
| 74 | 0.09898 | 0.1349 |
| 75 | 0.10289 | 0.1356 |
| 76 | 0.10680 | 0.1364 |
| 77 | 0.11072 | 0.1370 |
| 78 | 0.11465 | 0.1377 |
| 79 | 0.11858 | 0.1382 |
| 80 | 0.12253 | 0.1387 |
| 81 | 0.12648 | 0.1392 |
| 82 | 0.13043 | 0.1396 |
| 83 | 0.13440 | 0.1399 |
| 84 | 0.13837 | 0.1402 |
| 85 | 0.14234 | 0.1403 |
| 86 | 0.14632 | 0.1404 |
| 87 | 0.14950 | 0.1404 |
| 88 | 0.15190 | 0.1404 |
| 89 | 0.15430 | 0.1403 |
| 90 | 0.15670 | 0.1401 |
| 91 | 0.15909 | 0.1400 |
| 92 | 0.16149 | 0.1398 |
| 93 | 0.16388 | 0.1396 |
| 94 | 0.16628 | 0.1394 |
| 95 | 0.16867 | 0.1392 |
| 96 | 0.17106 | 0.1390 |
| 97 | 0.17345 | 0.1387 |
| 98 | 0.17584 | 0.1385 |
| 99 | 0.17823 | 0.1382 |
| 100 | 0.18062 | 0.1380 |
| 101 | 0.18301 | 0.1377 |
| 102 | 0.18539 | 0.1375 |
| 103 | 0.18778 | 0.1372 |
| 104 | 0.19017 | 0.1370 |
| 105 | 0.19255 | 0.1367 |
| 106 | 0.19494 | 0.1364 |
| 107 | 0.19733 | 0.1361 |
| 108 | 0.19971 | 0.1359 |
| 109 | 0.20210 | 0.1356 |
| 110 | 0.20448 | 0.1353 |
| 111 | 0.20687 | 0.1350 |
| 112 | 0.20925 | 0.1347 |
| 113 | 0.21163 | 0.1344 |
| 114 | 0.21402 | 0.1341 |
| 115 | 0.21640 | 0.1338 |
| 116 | 0.21879 | 0.1335 |
| 117 | 0.22117 | 0.1332 |
| 118 | 0.22355 | 0.1329 |
| 119 | 0.22593 | 0.1326 |
| 120 | 0.22832 | 0.1323 |
| 121 | 0.23070 | 0.1320 |
| 122 | 0.23308 | 0.1317 |
| 123 | 0.23546 | 0.1314 |
| 124 | 0.23785 | 0.1311 |
| 125 | 0.24023 | 0.1308 |
| 126 | 0.24261 | 0.1305 |
| 127 | 0.24499 | 0.1302 |
| 128 | 0.24737 | 0.1299 |
| 129 | 0.24975 | 0.1296 |
| 130 | 0.25213 | 0.1292 |
| 131 | 0.25452 | 0.1289 |
| 132 | 0.25690 | 0.1286 |
| 133 | 0.25928 | 0.1283 |
| 134 | 0.26166 | 0.1280 |
| 135 | 0.26404 | 0.1277 |
| 136 | 0.26642 | 0.1274 |
| 137 | 0.26880 | 0.1271 |
| 138 | 0.27119 | 0.1268 |
| 139 | 0.27357 | 0.1264 |
| 140 | 0.27595 | 0.1261 |
| 141 | 0.27833 | 0.1258 |
| 142 | 0.28071 | 0.1255 |
| 143 | 0.28310 | 0.1252 |
| 144 | 0.28548 | 0.1249 |
| 145 | 0.28786 | 0.1246 |
| 146 | 0.29025 | 0.1243 |
| 147 | 0.29263 | 0.1240 |
| 148 | 0.29501 | 0.1237 |
| 149 | 0.29740 | 0.1235 |
| 150 | 0.29978 | 0.1232 |
| 151 | 0.30217 | 0.1229 |
| 152 | 0.30456 | 0.1227 |
| 153 | 0.30695 | 0.1224 |
| 154 | 0.30934 | 0.1222 |
| 155 | 0.31173 | 0.1219 |
| 156 | 0.31412 | 0.1217 |
| 157 | 0.31651 | 0.1215 |
| 158 | 0.31891 | 0.1213 |
| 159 | 0.32130 | 0.1212 |
| 160 | 0.32370 | 0.1210 |
| 161 | 0.32610 | 0.1208 |
| 162 | 0.32849 | 0.1207 |
| 163 | 0.33089 | 0.1205 |
| 164 | 0.33329 | 0.1204 |
| 165 | 0.33568 | 0.1203 |
| 166 | 0.33808 | 0.1201 |
| 167 | 0.34000 | 0.1200 |
| 168 | 0.34205 | 0.1199 |
| 169 | 0.34409 | 0.1198 |
| 170 | 0.34614 | 0.1196 |
| 171 | 0.34818 | 0.1194 |
| 172 | 0.35022 | 0.1193 |
| 173 | 0.35226 | 0.1191 |
| 174 | 0.35431 | 0.1190 |
| 175 | 0.35635 | 0.1189 |
| 176 | 0.35840 | 0.1187 |
| 177 | 0.36045 | 0.1187 |
| 178 | 0.36250 | 0.1187 |
| 179 | 0.36455 | 0.1187 |
| 180 | 0.36659 | 0.1189 |
| 181 | 0.36864 | 0.1190 |
| 182 | 0.37068 | 0.1192 |
| 183 | 0.37272 | 0.1194 |
| 184 | 0.37475 | 0.1196 |
| 185 | 0.37679 | 0.1199 |
| 186 | 0.37883 | 0.1201 |
| 187 | 0.38087 | 0.1202 |
| 188 | 0.38292 | 0.1203 |
| 189 | 0.38497 | 0.1204 |
| 190 | 0.38702 | 0.1203 |
| 191 | 0.38906 | 0.1202 |
| 192 | 0.39110 | 0.1199 |
| 193 | 0.39312 | 0.1197 |
| 194 | 0.39514 | 0.1193 |
| 195 | 0.39714 | 0.1188 |
| 196 | 0.39913 | 0.1183 |
| 197 | 0.40110 | 0.1178 |
| 198 | 0.40309 | 0.1173 |
| 199 | 0.40506 | 0.1167 |
| 200 | 0.40705 | 0.1162 |
| 201 | 0.40906 | 0.1158 |
| 202 | 0.41108 | 0.1155 |
| 203 | 0.41312 | 0.1153 |
| 204 | 0.41517 | 0.1152 |
| 205 | 0.41721 | 0.1153 |
| 206 | 0.41925 | 0.1155 |
| 207 | 0.42127 | 0.1159 |
| 208 | 0.42325 | 0.1164 |
| 209 | 0.42521 | 0.1170 |
| 210 | 0.42715 | 0.1177 |
| 211 | 0.42906 | 0.1184 |
| 212 | 0.43098 | 0.1191 |
| 213 | 0.43290 | 0.1198 |
| 214 | 0.43483 | 0.1205 |
| 215 | 0.43680 | 0.1211 |
| 216 | 0.43879 | 0.1216 |
| 217 | 0.44082 | 0.1218 |
| 218 | 0.44287 | 0.1217 |
| 219 | 0.44490 | 0.1215 |
| 220 | 0.44690 | 0.1211 |
| 221 | 0.44888 | 0.1205 |
| 222 | 0.45080 | 0.1198 |
| 223 | 0.45271 | 0.1191 |
| 224 | 0.45459 | 0.1182 |
| 225 | 0.45646 | 0.1174 |
| 226 | 0.45832 | 0.1165 |
| 227 | 0.46017 | 0.1157 |
| 228 | 0.46205 | 0.1149 |
| 229 | 0.46395 | 0.1141 |
| 230 | 0.46590 | 0.1134 |
| 231 | 0.46787 | 0.1129 |
| 232 | 0.46988 | 0.1125 |
| 233 | 0.47191 | 0.1122 |
| 234 | 0.47396 | 0.1122 |
| 235 | 0.47601 | 0.1123 |
| 236 | 0.47804 | 0.1125 |
| 237 | 0.48006 | 0.1129 |
| 238 | 0.48206 | 0.1133 |
| 239 | 0.48404 | 0.1138 |
| 240 | 0.48601 | 0.1144 |
| 241 | 0.48798 | 0.1150 |
| 242 | 0.48995 | 0.1155 |
| 243 | 0.49194 | 0.1160 |
| 244 | 0.49394 | 0.1165 |

| 245 | 0.49596 | 0.1168 |
| --- | --- | --- |
| 246 | 0.49800 | 0.1170 |
| 247 | 0.50005 | 0.1171 |
| 248 | 0.50210 | 0.1170 |
| 249 | 0.50413 | 0.1168 |
| 250 | 0.50615 | 0.1165 |
| 251 | 0.50815 | 0.1160 |
| 252 | 0.51013 | 0.1155 |
| 253 | 0.51208 | 0.1148 |
| 254 | 0.51402 | 0.1142 |
| 255 | 0.51595 | 0.1135 |
| 256 | 0.51784 | 0.1127 |
| 257 | 0.51972 | 0.1119 |
| 258 | 0.52163 | 0.1111 |
| 259 | 0.52356 | 0.1105 |
| 260 | 0.52552 | 0.1099 |
| 261 | 0.52752 | 0.1094 |
| 262 | 0.52954 | 0.1091 |
| 263 | 0.53158 | 0.1089 |
| 264 | 0.53363 | 0.1088 |
| 265 | 0.53567 | 0.1089 |
| 266 | 0.53771 | 0.1091 |

## A5 Kulite location

| Number | Position (z/L) |
| --- | --- |
| 1 | 0.129 |
| 2 | 0.596 |
| 3 | 0.68 |
| 4 | 0.72 |
| 5 | 0.735 |
| 6 | 0.766 |
| 7 | 0.81 |